\documentclass{emulateapj}
\usepackage{apjfonts}

\usepackage{lscape}
\usepackage{graphicx}
\usepackage{natbib}
\usepackage{longtable}

\newcommand{\etal}{et al.}

\newcommand{\CIV}{C{\sevenrm IV}}

\def\MgII{Mg\,{\sc ii}}

\newcommand{\MgIIab}{Mg{\sevenrm II}\,$\lambda\lambda$2796,2803}
\newcommand{\OIIwave}{[O{\sevenrm\,II}]\,$\lambda$3727}
\newcommand{\OII}{[O{\sevenrm\,II}]}
\newcommand{\NeVwave}{[Ne{\sevenrm\,V}]\,$\lambda\lambda$3346,3426}

\def \OIII {[O\,{\sc iii}]}

 \font\sevenrm=cmr7 scaled 1000

\begin{document}

\title{On the link between associated \MgII\ absorbers and star formation in quasar hosts}

\shorttitle{QUASARS WITH MGII AALS}


\shortauthors{SHEN \& M\'{E}NARD}
\author{Yue Shen\altaffilmark{1} and Brice M\'{e}nard\altaffilmark{2,3,4} }
\altaffiltext{1}{Harvard-Smithsonian Center for Astrophysics, 60
Garden St., MS-51, Cambridge, MA 02138, USA; yshen@cfa.harvard.edu}
\altaffiltext{2}{CITA, University of Toronto, 60 St. George Street, Toronto, Ontario, M5S 3H8, Canada}
\altaffiltext{3}{Institute for Cosmic Ray Research, University of Tokyo, Kashiwa 2778582, Japan}
\altaffiltext{4}{Department of Physics and Astronomy, Johns Hopkins University, Baltimore, MD 21218, USA; menard@pha.jhu.edu}

\begin{abstract}
A few percent of quasars show strong associated \MgII\ absorption, with
velocities ($v_{\rm off}$) lying within a few thousand ${\rm km\,s^{-1}}$
from the quasar systemic redshift. These associated absorption line systems
(AALs) are usually interpreted as absorbers that are either intrinsic to the
quasar and its host, or arising from external galaxies clustering around the
quasar. Using composite spectra of $\sim 1,800$ \MgII\ AAL quasars selected
from SDSS DR7 at $0.4\la z\la 2$, we show that quasars with AALs with $v_{\rm
off}<1500\, {\rm km\,s^{-1}}$ have a prominent excess in \OIIwave\ emission
(detected at $>7\sigma$) at rest relative to the quasar host, compared to
unabsorbed quasars. We interpret this \OII\ excess as due to enhanced star
formation in the quasar host. Our results suggest that a significant fraction
of AALs with $v_{\rm off}<1500\,{\rm km\,s^{-1}}$ are physically associated
with the quasar and its host. AAL quasars also have dust reddening lying
between normal quasars and the so-called dust-reddened quasars. We suggest
that the unique properties of AAL quasars can be explained if they are the
transitional population from heavily dust-reddened quasar to normal quasars
in the formation process of quasars and their hosts. This scenario predicts a
larger fraction of young bulges, disturbed morphologies and interactions of
AAL quasar hosts compared to normal quasars. The intrinsic link between
associated absorbers and quasar hosts opens a new window to probe massive
galaxy formation and galactic-scale feedback processes, and provides a
crucial test of the evolutionary picture of quasars.
\end{abstract}

\section{Introduction}

In the popular Sanders \& Mirabel picture for quasar evolution
\citep[e.g.,][]{Sanders_etal_1988,Sanders_Mirabel_1996,Canalizo_Stockton_2001,
Hopkins_etal_2008}, gas-rich major mergers between galaxies trigger intense
starbursts in {ultraluminous infrared galaxies} (ULIRGs) and channel ample
fuel into the nuclear region to feed the black hole
\citep[e.g.,][]{Hernquist_1989}. The black hole (BH) growth is initially
buried in dust until some feedback process (either stellar-driven or
quasar-driven) blows the dust away and a classical quasar emerges. The major
merger event also transforms the host morphology from disks to a spheroid
\citep[e.g.,][]{Toomre_Toomre_1972}. The feedback processes invoked in this
model are also thought to be responsible for shutting down star formation in
the bulge, and for establishing the tight correlations between BH mass and
bulge properties as observed in local dormant massive galaxies
\citep[e.g.,][]{Tremaine_etal_2002}. Although such a violent route is not
required for triggering low-to-intermediate luminosity Active Galactic
Nucleus (AGN) activity, this merger-based framework seems successful in
reproducing an array of observations of luminous quasars and massive red
ellipticals in the cosmological context
\citep[e.g.,][]{Wyithe_Loeb_2003,Hopkins_etal_2008,Shen_2009}. Nevertheless,
many fundamental issues remain unresolved in this model, in particular, the
nature and physics of these fueling and feedback processes along the
evolutionary sequence remain largely elusive.

A useful tool to probe the physical conditions and dynamical processes within
quasar hosts is intrinsic quasar absorption lines. Historically most of the
focus has been on broad absorption lines (BALs, usually defined as absorption
troughs broader than $2000\,{\rm km\,s^{-1}}$), which are undoubtedly
intrinsic to the quasar. Here we focus on another class of narrow absorption
lines called associated absorption lines (AALs), whose absorption velocity is
close to the systemic velocity of the background quasar ($z_{ab}\approx
z_{em}$). AALs are traditionally defined as narrow absorption troughs ($\la
500\,{\rm km\,s^{-1}}$) with a velocity offset $v_{\rm off}$ within $\pm
3000\,{\rm km\,s^{-1}}$ of the systemic redshift of the quasar \footnote{Some
earlier studies used a larger velocity offset cut $\sim 5000-6000\, {\rm
km\,s^{-1}}$ to define AALs. This is partly due to the fact that the systemic
redshift for many high-redshift quasars is based on the centroid of broad
emission lines such as \CIV, which has a larger uncertainty compared with the
systemic redshift determined from \MgII\ or narrow emission lines.}. Strong
(${\rm EW}>0.6\,$\AA) low-ionization \MgII\ associated absorbers are present
in a few percent of the entire quasar population
\citep[e.g.,][]{Vanden_Berk_etal_2008}. These absorption systems are
generally believed to be close to the quasar and are explained by either (or
a combination) of the following scenarios: absorption by external galaxies
clustering around the quasar
\citep[e.g.,][]{Weymann_etal_1979,Wild_etal_2008}; absorption by halo clouds
of the quasar host galaxy \citep[e.g.,][]{Heckman_etal_1991}; absorption by
material from a starburst wind of the quasar host
\citep[e.g.,][]{Heckman_etal_1990}; or originating from the vicinity of the
black hole ($\la 200$\,pc) based on partial coverage analysis or variability
studies \citep[e.g.,][]{Hamann_etal_1995,Barlow_Sargent_1997}. On the other
hand, classical intervening absorber systems (with $z_{ab}\ll z_{em}$) are
not physically associated with the quasar and are absorptions due to
cosmologically intervening foreground galaxies along the quasar line-of-sight
(LOS) \citep{1969ApJ...156L..63B,1986A&A...155L...8B}. The strength of these
intervening absorbers is shown to correlate with the associated star
formation rate, measured within 10 kpc
\citep{Menard_etal_2011}.

Using composite spectra, \citet{Vanden_Berk_etal_2008} showed that AAL
quasars on average have dust extinction almost twice that observed in quasars
with intervening absorbers \citep[e.g.,][]{York_etal_2006}, and the
associated absorbers have a higher ionization state than intervening
absorbers. To some extent, this supports the idea that AALs are physically
associated with the quasar (or at least affected by the radiation field of
the quasar), but a direct link between AALs and the quasar is still elusive
\citep[cf.,][]{Wild_etal_2008}.

Using $\sim 1800$ quasars with associated \MgII\ absorbers from the SDSS, we
will show that there is a link between AALs and \OII\ emission from the
quasar host, suggesting that a substantial fraction of AALs are intrinsic to
the quasar and its host. We further argue that the properties of AAL quasars
are in favor of them being the transitional population in the Sanders \&
Mirabel picture, with signatures of feedback. We describe our sample in
\S\ref{sec:data} and present our main results in \S\ref{sec:composite}. In
\S\ref{sec:disc} we present an evolutionary picture to interpret our results
and discuss its implications.

\section{Data}\label{sec:data}

\begin{figure}
 \centering
 \includegraphics[width=0.48\textwidth]{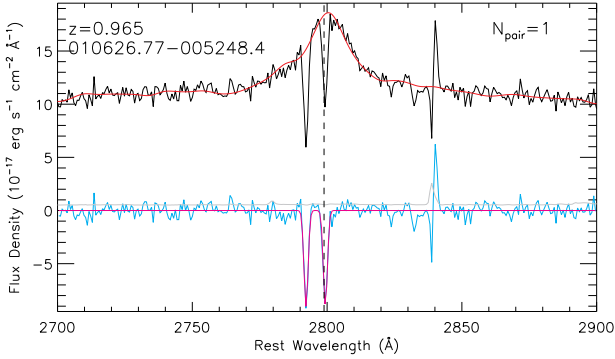}
    \caption{An example of our fitting procedure to identify \MgII\ AALs. The black
    and gray lines show the SDSS spectrum and flux density errors around the \MgII\
    region. The red line shows our pseudo-continuum plus emission line model fit.
    The dashed line marks the quasar systemic velocity based on the centroid of
    the broad \MgII\ line. The cyan line shows the continuum+emission line subtracted
    spectrum and the magenta line shows the double-Gaussian fit to the absorption
    doublet. For this example only $N_{\rm pair}=1$ absorption doublet was identified.}\label{fig:aas_examp}
\end{figure}

\begin{figure}
 \centering
 \includegraphics[width=0.48\textwidth]{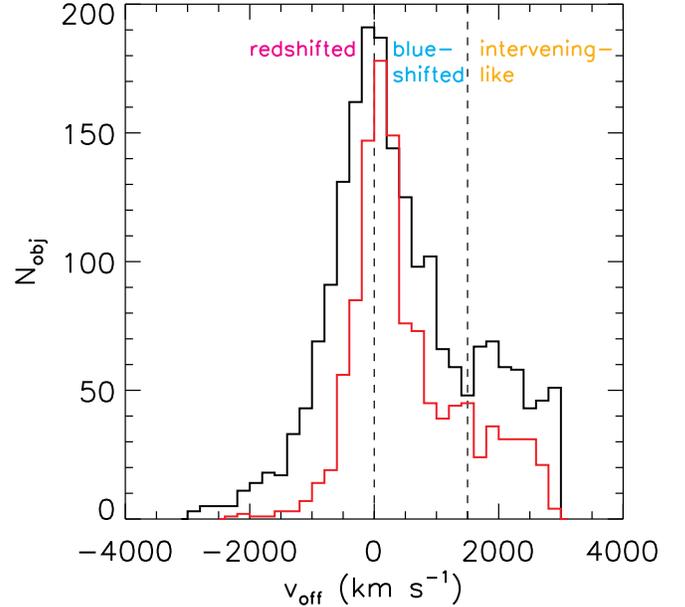}
    \caption{
    Distribution of velocity offset $v_{\rm off}$ of associated \MgII\ absorbers. Positive (negative) values
    indicate blueshifted (redshifted) from the systemic redshift. The black histogram shows the whole sample while the red one represents a subset of
    quasars with $z<1.4$ for which we have emission line redshifts based on \OII\ or \OIII. Systems with $v_{\rm off}> 1500\ {\rm km\,s^{-1}}$ are
    found to be similar to classical intervening absorber systems.
}\label{fig:vdist}
\end{figure}

\begin{figure}[]
 \centering
 \includegraphics[width=0.48\textwidth]{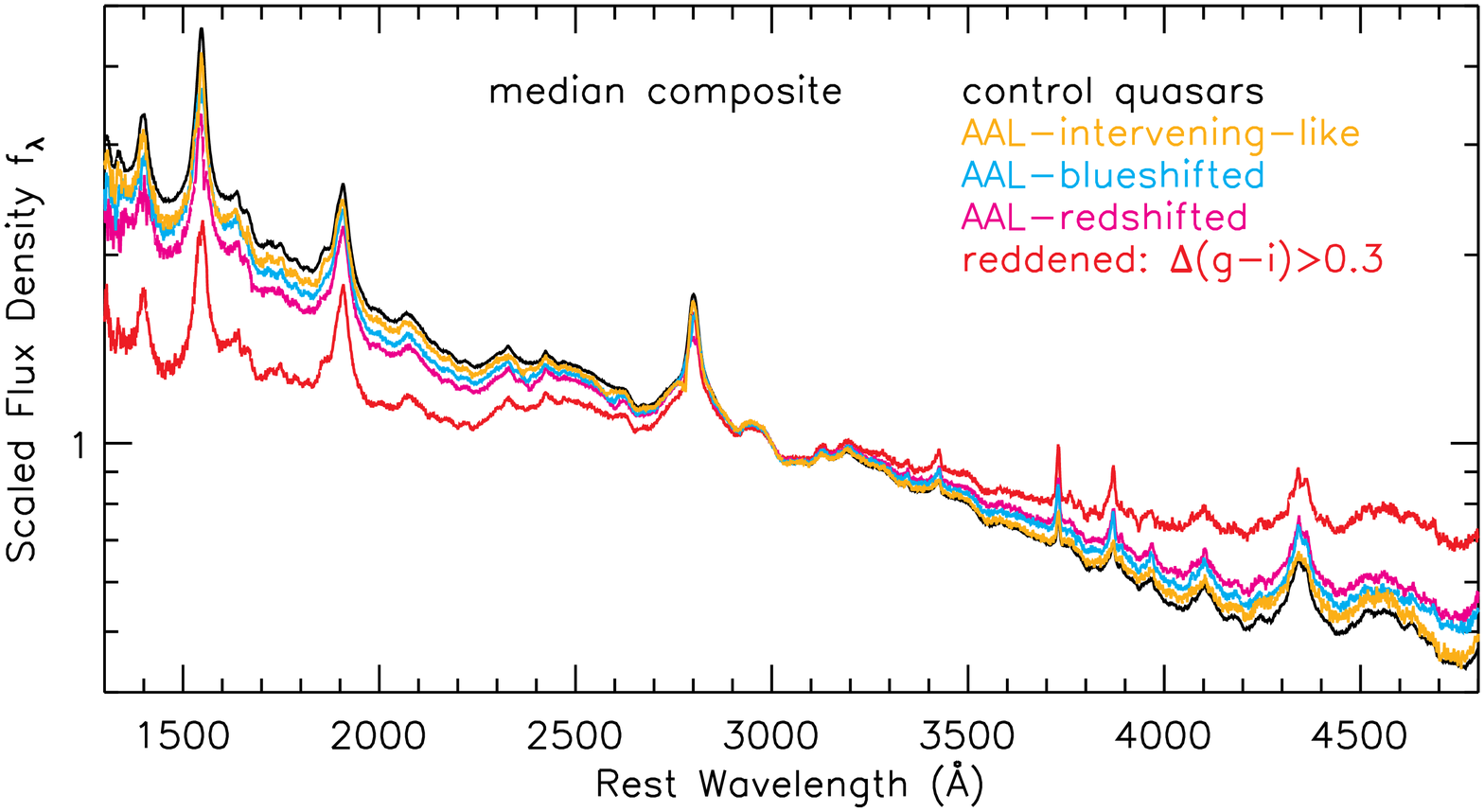}
 \includegraphics[width=0.48\textwidth]{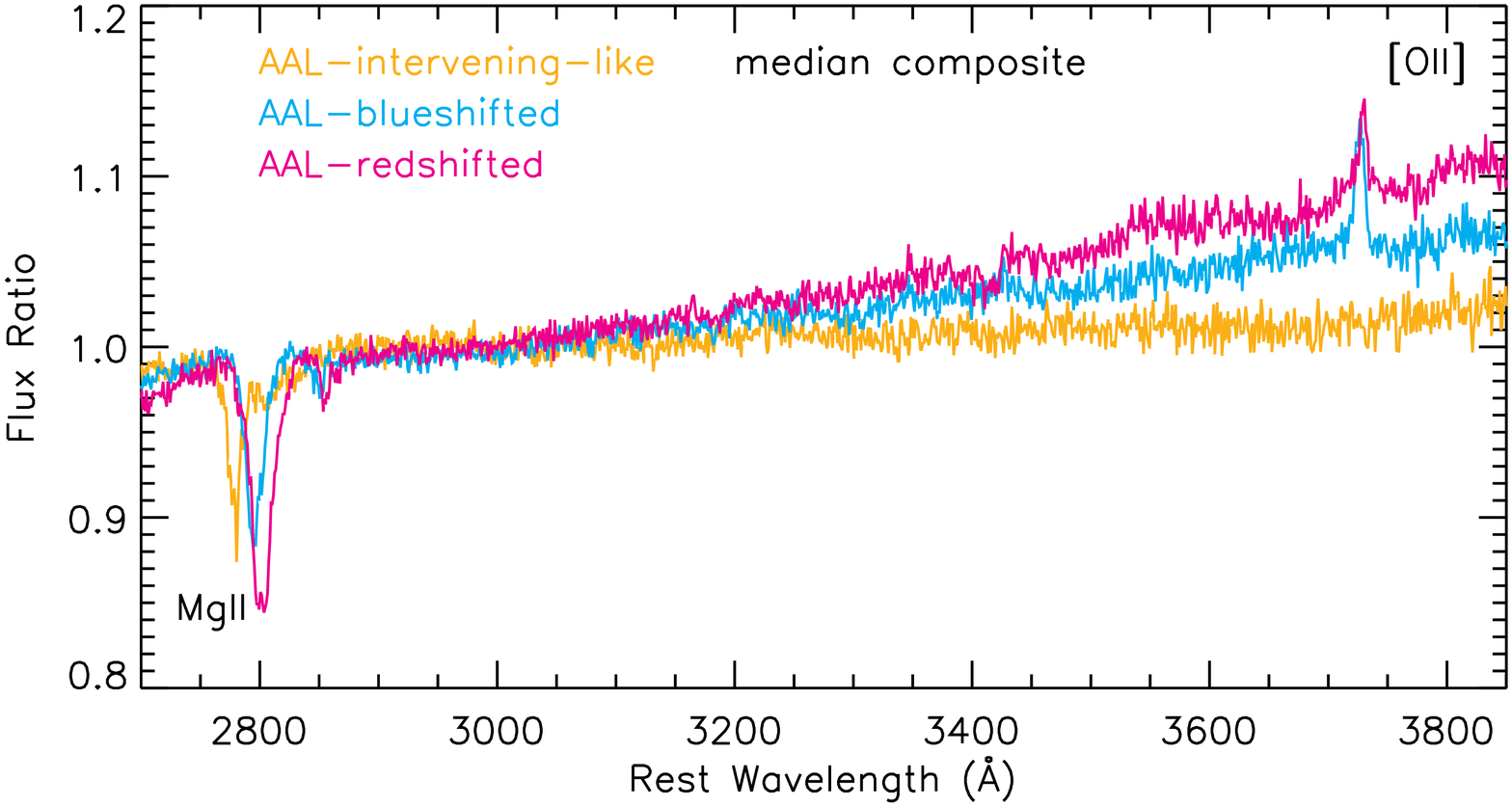}
    \caption{{\em Top:} Median composite spectra for different samples, stacked in the quasar rest frame and  normalized at 3000\,\AA.
    The three AAL samples (intervening-like:yellow, blueshifted:cyan, and redshifted:magenta)
    show increasing reddening relative to the control sample (black), but are substantially less reddened
than the SDSS dust-reddened quasars (with color excess $\Delta(g-i)>0.3$)
shown in red. {\em Bottom:} Flux ratios between composite spectra of the three
AAL samples and the control normal quasars. While the AAL samples show (by
construction) \MgII\ absorption with different velocities, an excess of \OII\
emission is seen for the blueshifted and redshifted samples {\em at the
quasar systemic redshift}. }\label{fig:composite}
\end{figure}

\begin{deluxetable*}{lccccccccccc}
\tablecaption{The Sample of AALs \label{table:sample}} \tablehead{ SDSS
Designation & Plate & Fiber & MJD & redshift & EW$_{2796}$ &EW Err$_{2796}$ &
EW$_{2803}$ & EW Err$_{2803}$ & $v_{\rm off}$ & $v_{\rm off,SDSS}$ & $v_{\rm
off,HW}$ \\
(hhmmss.ss$+$ddmmss.s) & & & & & (\AA) &(\AA) &(\AA) &(\AA) & $({\rm
km\,s^{-1}})$ & $({\rm km\,s^{-1}})$ & $({\rm km\,s^{-1}})$ } \startdata
000045.77$+$255106.1 & 2822 & 339 & 54389 & $1.4446$ & $0.61$ & 0.07 & $0.50$ & 0.07 & 1300 & 2726 & 2069\\
000140.70$+$260425.5 & 2822 & 322 & 54389 & $0.7653$ & $0.48$ & 0.04 & $0.47$ & 0.04 & 47   & 54   & 113\\
000219.64$+$260029.2 & 2822 & 367 & 54389 & $1.2507$ & $3.07$ & 0.18 & $3.00$ & 0.14 & -421 & -588 & -289\\
\enddata
\tablecomments{The list of AALs used in this paper. In the last
three columns we list the absorber velocity offset calculated from the
fiducial systemic velocity (based on the broad \MgII\ centroid), the SDSS
redshift in the DR7 quasar catalog, and the improved redshift from
\citet{Hewett_Wild_2011}. The full catalog is available in the electronic version.}
\end{deluxetable*}

We search all the DR7 quasars \citep{Schneider_etal_2010} with \MgII\
coverage ($\sim 85$\,k quasars) for associated \MgII\ absorption. This
restricts our sample to $0.4\la z\la 2$ ($\bar{z}\sim 1.2$), which extends to
lower redshift than the sample in \citet{Vanden_Berk_etal_2008}. The
unabsorbed continuum plus emission line flux is modeled with a $\chi^2$ fit
with rejections of absorption troughs \citep{Shen_etal_2011}. We refer the
reader to that paper on the details of our fitting procedure. In short, we
fit the restframe 2200-3090\ \AA\ region with a power-law continuum plus an
iron template, and a set of Gaussians for the \MgII\ line. During the
fitting, we mask out 3$\sigma$ pixel outliers below the 20 pixel
boxcar-smoothed spectrum to minimize the effects of narrow absorption
troughs. We then subtract the pseudo-continuum plus emission line emission
from the spectrum to get the residual spectrum. \MgIIab\ absorption doublets
are then identified on both sides of the \MgII\ emission line by fitting
double-Gaussians to the absorption doublet. We only keep absorbers that are
detected at $>3\sigma$ for both Gaussian components. We measure restframe
equivalent widths (EW) of these AALs and uncertainties of EWs are estimated
from the Gaussian fits. Fig.\ \ref{fig:aas_examp} shows an example of our
fitting results. Although this fitting recipe was developed to reduce the
effects of narrow absorptions on the emission line measurements rather than a
dedicated recipe for finding narrow absorption lines, it does a reasonably
good job in identifying AALs. With this method we recovered $\sim 80\%$ of
the AALs in the DR3 sample of \citet{Vanden_Berk_etal_2008}, where most of
the ``missing'' AALs are weak absorptions with otherwise normal quasar
properties. However, we do not intend to quantify the completeness in our AAL
selection as functions of S/N and absorber strength, as such a task would
require a more dedicated narrow absorption finder and Monte-Carlo
simulations, which are unnecessary for the purposes of this paper.

To define an AAL one needs to know the systemic redshift of the quasar.
Stellar absorption features are generally unavailable due to the overwhelming
quasar light. The redshifts based on narrow lines such as \OII\ and \OIII\
are generally consistent with stellar absorption redshifts within $\la
100\,{\rm km\,s^{-1}}$ \citep[e.g.,][]{Richards_etal_2002b}. The redshifts
based on the broad \MgII\ emission line are consistent with those based on
\OII\ or \OIII\ with a dispersion of $\sim 350\,{\rm km\,s^{-1}}$
\citep[e.g.,][their fig.\ 17]{Richards_etal_2002b,Shen_etal_2011}. The
high-ionization broad \CIV\ line is known to be systematically blueshifted
($\sim 700\,{\rm km\,s^{-1}}$) from low-ionization lines such as \MgII, with
a large dispersion of $\sim 700\,{\rm km\,s^{-1}}$
\citep[e.g.,][]{Gaskell_1982,Tytler_Fan_1992,Richards_etal_2002b,Shen_etal_2011},
and hence will bias the systemic redshift determination at high redshift.

Here we adopt the centroid of the broad \MgII\ emission line as the quasar
systemic redshift to calculate the absorber velocity offset, and use
$3000\,{\rm km\,s^{-1}}$ as the velocity cut to define an AAL. We compared
the absorber velocity offsets with those calculated by adopting the improved
redshifts of SDSS quasars from \citet{Hewett_Wild_2011}, and found good
agreements with a dispersion of $\sim 300\,{\rm km\,s^{-1}}$. This dispersion
is consistent with the redshift uncertainty $\sim 350\,{\rm km\,s^{-1}}$
based on the \MgII\ centroid. Therefore we adopt a nominal uncertainty of
$\Delta v_{\rm off}=350\,{\rm km\,s^{-1}}$ in absorber velocities, which is
dominated by systemic redshift uncertainties. We finally arrived at a sample
of $\sim 1800$ quasars with AALs, among which $\sim 10\%$ show multiple AALs
in SDSS spectra \citep[consistent with][]{Wild_etal_2008}. The sample of AALs used in
the following analysis is listed in Table \ref{table:sample}. Additional
information about these quasars can be retrieved from the value-added SDSS
DR7 quasar catalog compiled in \citet{Shen_etal_2011}.

Fig.\ \ref{fig:vdist} shows the distribution of absorber velocity $v_{\rm
off}$ for the whole sample in black, and for a subset of $z<1.4$ quasars for
which we have redshifts from \OII\ or \OIII\ in red. Positive values indicate
blueshifted relative to the systemic and negative values indicate redshifted.
We found that the absorber velocity distribution using this systemic redshift
definition is similar (albeit slightly broader due to the less accurate
\MgII\ emission line redshifts) to that using the subset of our sample with
\OII\ or \OIII-based redshifts. The dispersion in the $v_{\rm off}$
distribution is $\sim 600\,{\rm km\,s^{-1}}$, larger than the typical
uncertainty $\Delta v_{\rm off}\sim 350\,{\rm km\,s^{-1}}$ arising from
systemic redshift uncertainties based on \MgII. It is thus
 dominated by an intrinsic dispersion. Our composite spectra (see
\S\ref{sec:composite}) also confirm that this dispersion
is not caused by incorrect systemic redshifts.

As earlier studies already show
\citep[e.g.,][]{Weymann_etal_1979,Vanden_Berk_etal_2008}, there is a
substantial fraction of AALs that are redshifted from the systemic velocity.
These systems could either be truly infalling absorbers, or the emission
lines used to estimate the systemic redshift have significant outflowing
velocity. At $v_{\rm off}>1500\,{\rm km\,s^{-1}}$, the distribution flattens
out and is consistent with the expectation from cosmologically intervening
absorbers, as seen in earlier studies \citep[e.g.,][]{Wild_etal_2008}. We
thus argue that these high-velocity AALs are mostly classical intervening
absorbers, which will be confirmed by our composite spectra in
\S\ref{sec:composite}.

We divide our sample into three categories: 1) AAL quasars with
$v_{\rm off}>1500\,{\rm km\,s^{-1}}$ (the ``intervening-like''
sample); 2) AAL quasars with $0<v_{\rm off}<1500\,{\rm
km\,s^{-1}}$ (the ``blueshifted'' sample); 3) AAL quasars with
$-3000<v_{\rm off}<0 \,{\rm km\,s^{-1}}$ (the ``redshifted''
sample). We have $\sim 750$ quasars each in the latter two
samples, and $\sim 370$ quasars in the former.

\section{Composite Spectra}\label{sec:composite}


\begin{figure*}
 \centering
 \includegraphics[width=0.45\textwidth]{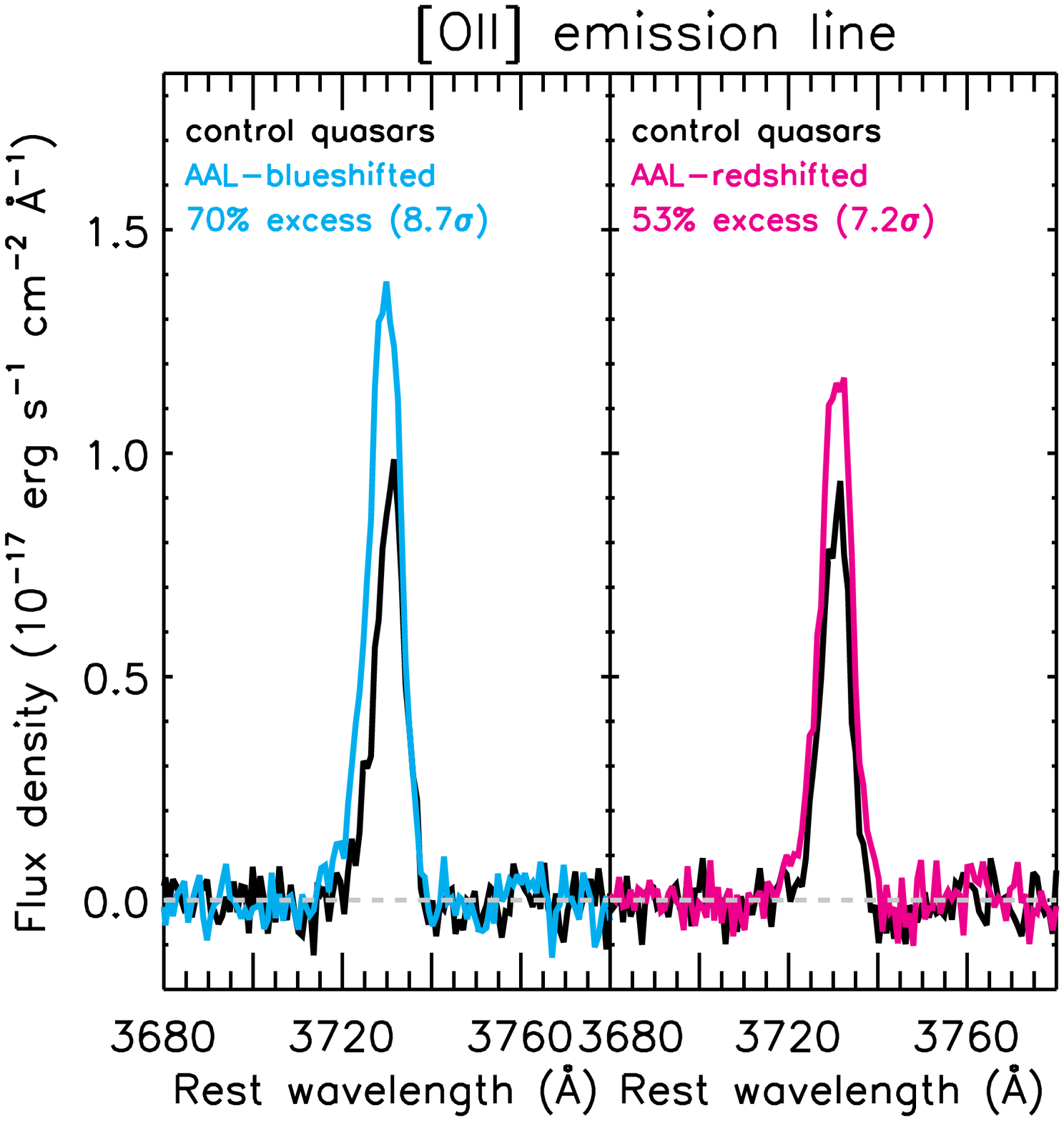}
	\hspace{.1cm}
 \includegraphics[width=0.45\textwidth]{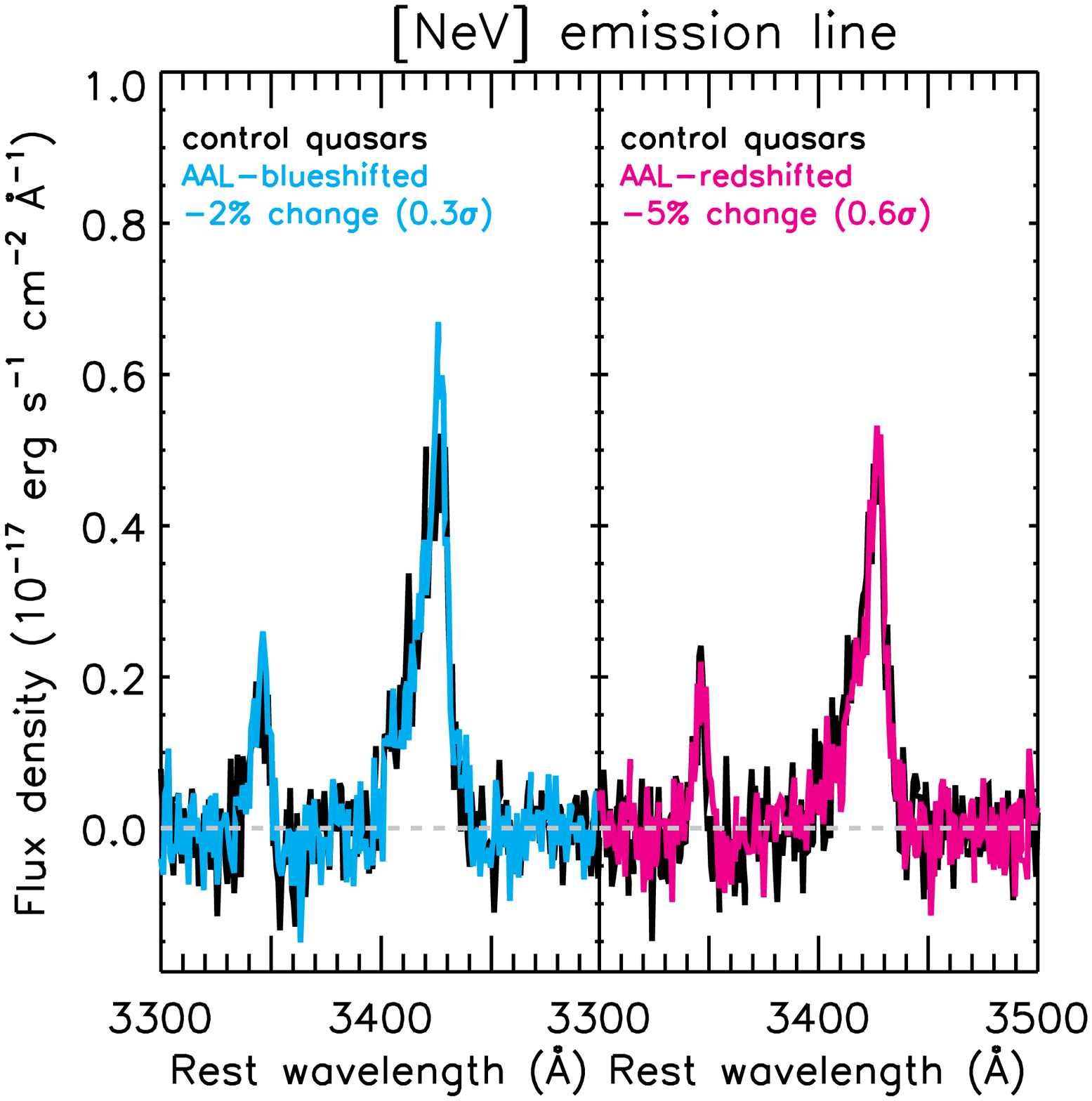}
    \caption{ \emph{Left:} Continuum-subtracted median composites of the \OII\ emission line, stacked in the quasar rest frame. The black lines show
    the level of emission for the control quasars. We detect a $70$ and $53\%$ excess for the blueshifted and redshifted samples respectively.
    \emph{Right:} Same quantify shown for the \NeVwave\ emission line. In this case we do not detect any change in the line flux.}
    \label{fig:line_composite}
\end{figure*}


To study the average properties of these three AAL quasar samples we create
high S/N composite spectra in the restframe of the quasar. SDSS spectra
integrate all the light received within a 1.5\arcsec\ fiber radius,
corresponding to $\sim 10$ kpc at $z\sim 1$, hence they cover a substantial
fraction of the quasar host. For comparison purposes we construct a control
sample of more than 9,000 quasars without detected AALs, matched in redshift
and $i$-band magnitude to the AAL quasars. We create composite spectra using
the method described in \citet[][]{VandenBerk_etal_2001}. In short, each
spectrum was shifted to restframe, rebinned onto a common dispersion of 1
\AA\ per bin, and normalized. The final composite spectrum was generated by
taking the median (or geometric mean) flux density in each bin of the
shifted, rebinned, and scaled spectra. An error spectrum was generated from
the 68\% semi-interquantile range of the flux densities in each bin scaled by
$N_{\rm spec}^{1/2}$, where $N_{\rm spec}$ is the number of spectra
contributing to that bin. We refer the reader to \citet{VandenBerk_etal_2001}
for more details regarding generating the composite spectra.

The left panel of Fig.\ 3 shows {\em median} composite spectra for the above
quasar samples. AAL quasars have colors lying between normal quasars and the
so-called ``dust-reddened'' quasars in SDSS (with color
excess\footnote{$\Delta(g-i)$ is a good indicator of intrinsic quasar colors
which accounts for the band shifting with redshift
\citep[e.g.,][]{Richards_etal_2003}; larger $\Delta(g-i)$ values indicate
redder colors.} $\Delta(g-i)>0.3$; Richards et al. 2003). The reddening of
AAL quasars relative to the control quasars is well described by a SMC-like
extinction curve, with $E(B-V)\sim 0.03$, consistent with
\cite{Vanden_Berk_etal_2008}. The right panel shows flux ratios of the AAL
quasar composites to that of the control quasars. The ``redshifted'' and
``blueshifted'' samples show a prominent excess of narrow \OII\ emission,
which is absent in the case of the ``intervening-like'' sample. The lack of
\OII\ emission excess, the similar amount of reddening to classical
intervening absorption systems (e.g., York et al.\ 2006), and the fact that
they join the plateau of the velocity distribution (e.g., Fig.\
\ref{fig:vdist}), suggest that AALs with $v_{\rm off}>1500\,{\rm km\,s^{-1}}$
are in fact mostly classical intervening absorbers\footnote{This statement is
for the low-$z$ \MgII\ AALs, and may not hold true for the high-$z$ and
high-ionization AALs (such as \CIV). }. We note that the large-scale
fluctuations seen in the composite ratios are due to correlated noise arising
from variance in the shape of quasar continua.

The correlation between the presence of absorbers with $v_{\rm off}<1500$
km\,s$^{-1}$ and \OII\ emission is of great interest. While the composite
\MgII\ absorption lines are, by construction, offset by hundreds of ${\rm
km\,s^{-1}}$ with respect to the quasars, the stacked excess \OII\ emission
is found more or less at the quasar systemic velocity (see below). \emph{Thus
the excess \OII\ emission originates from material lying at the quasar
systemic velocity}. The central engine, i.e., the black hole radiation,
appears to be similar in AAL quasars and in normal quasars: other than the
reddening and \MgII\ absorption, no difference is seen in the continuum and
broad emission lines between the two quasar populations. This suggests that
dust-reddening is the explanation for the color difference seen in AAL
quasars and normal quasars. One might ask if this dust reddening might also
cause the apparent \OII\ excess seen in the flux ratio plot in
Fig.~\ref{fig:composite}. Dust reddening could occur on spatial scales much
smaller than the \OII\ emission region but much larger than the continuum
plus broad line region. In this case the continuum is attenuated while the
\OII\ emission is not, causing an apparent enhancement of \OII\ strength
relative to the underlying continuum. Assuming an SMC-like extinction curve,
we require $E(B-V)\sim 0.12$ to achieve the level of \OII\ emission
enhancement relative to the continuum, substantially larger than the inferred
$E(B-V)\sim 0.03$ from the composite spectra.

We now focus on the \OII\ emission excess. To quantify it and estimate its
significance, we create continuum-subtracted median composites around the
\OII\ emission line. This is done by subtracting a running median filter with
a width of 20 pixels and by masking the region $3710<\lambda<3750$ \AA\ to
preserve the emission line. Within this region the continuum estimate is
interpolated. The results are shown in Fig.~\ref{fig:line_composite}. The
centroid of the excess \OII\ emission is within $\sim 150\,{\rm km\,s^{-1}}$
from the systemic velocity (Fig.\ \ref{fig:line_composite}). Considering the
typical redshift uncertainties of individual objects used in the stacking
this is consistent with the systemic velocity. We measure the line flux by
integrating the flux density over the emission line profile. The line flux
error is estimated by bootstrapping the sample of reference quasars and
repeating the procedure 100 times. This allows us to include sample variance
in the error estimation. The \OII\ excess is detected at 8$\sigma$
(7$\sigma$) for the blueshifted (redshifted) sample, corresponding to an
enhancement of $70\%$ ($53\%$) of the line flux of the control quasars. The
enhancement increases to $94\%$ ($63\%)$ if we use mean composites. For
comparison, we measure the properties of the \NeVwave, which is entirely
excited by AGN given its high ionization potential. We follow the same
procedure and the results are presented in the right panel of
Fig.~\ref{fig:line_composite}. In this case, we do not observe any
significant excess in the \NeVwave. Therefore the AGN ionizing source of AAL
quasars is similar to that of normal quasars, and we conclude that the excess
\OII\ emission is due to different host properties. Interestingly, no \OII\
excess is seen in the composite spectrum of $\sim 300$ quasars with broad
\MgII\ absorption lines selected from SDSS \citep{Shen_etal_2011}, which is
expected if broad absorption lines (BALs) are ubiquitous among quasars and
the appearance of BALQSOs only depends on LOS. This difference reinforces our
conclusion that the hosts of AAL quasars are intrinsically different.

\section{Discussion}\label{sec:disc}

\begin{figure*}[!t]
 \centering
 \includegraphics[width=0.3\textwidth]{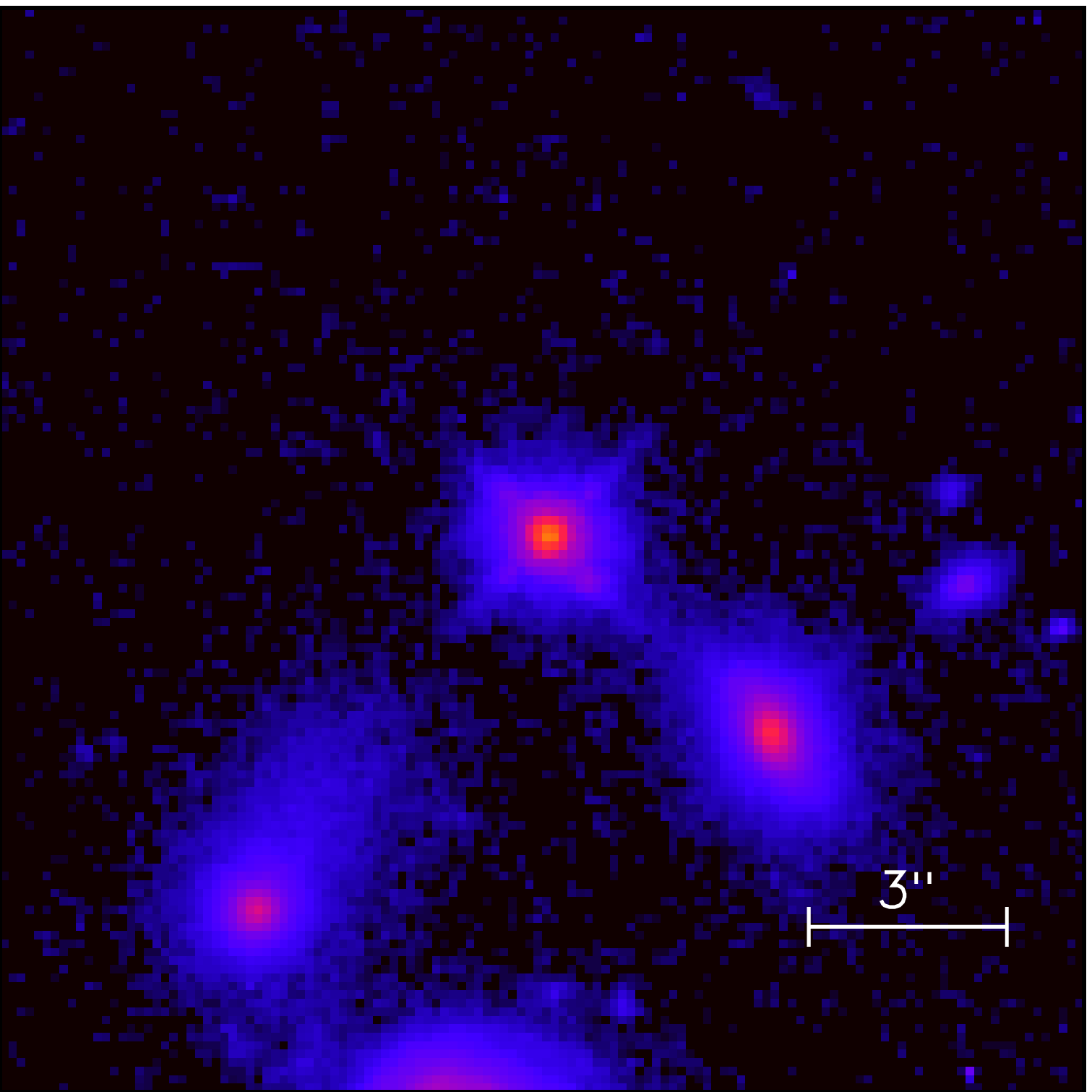} 
 \includegraphics[width=0.3\textwidth]{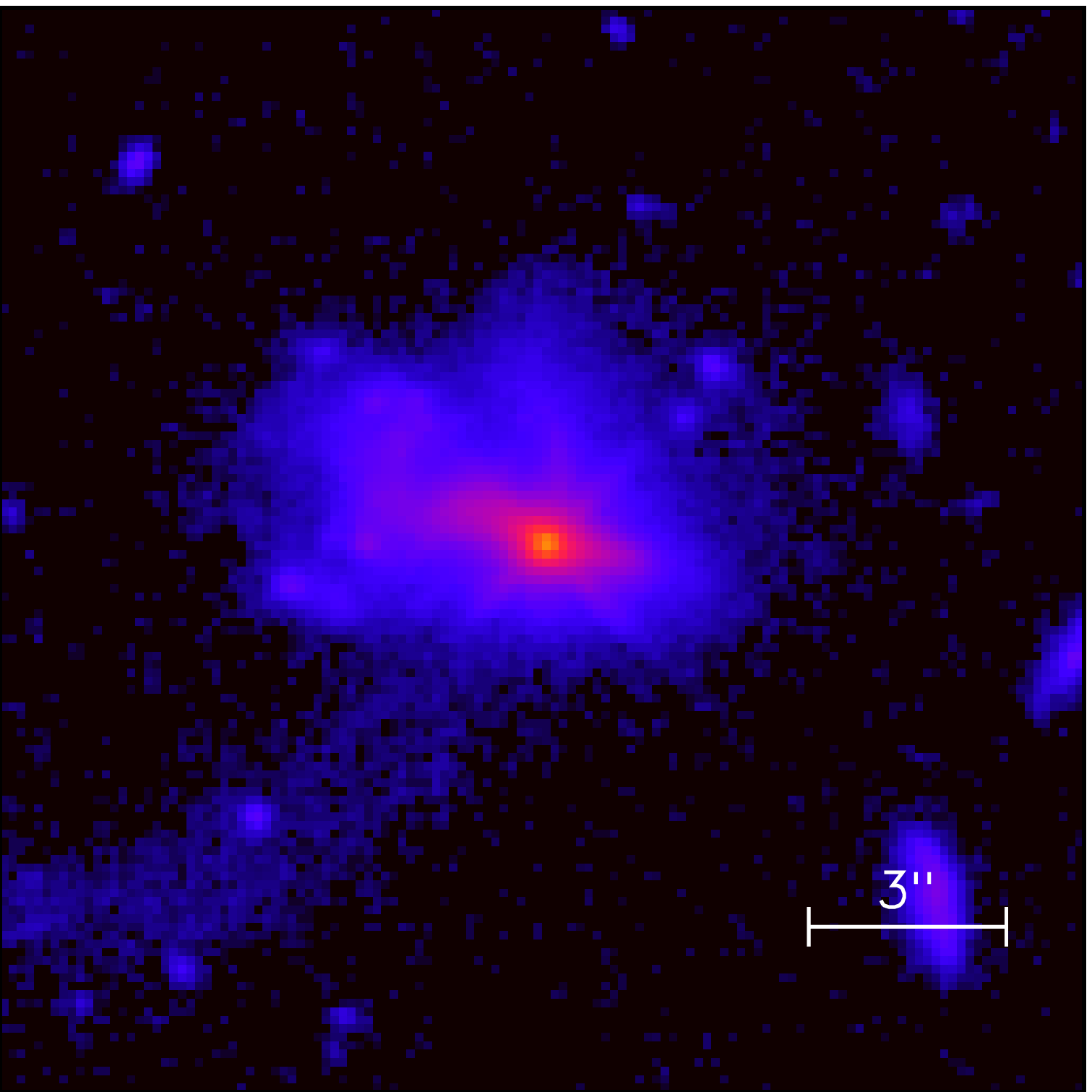} 
 \includegraphics[width=0.3\textwidth]{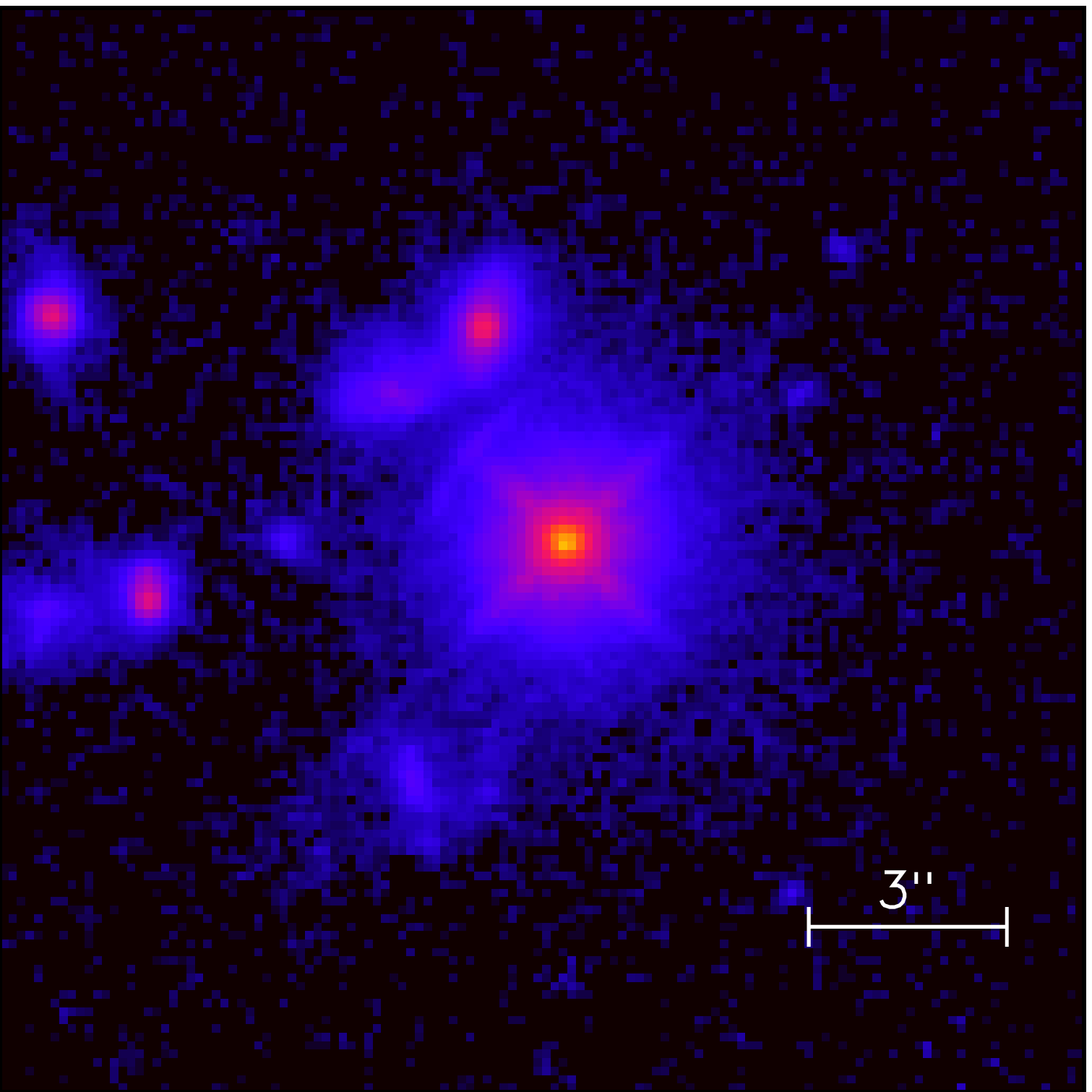}
 \includegraphics[width=0.3\textwidth]{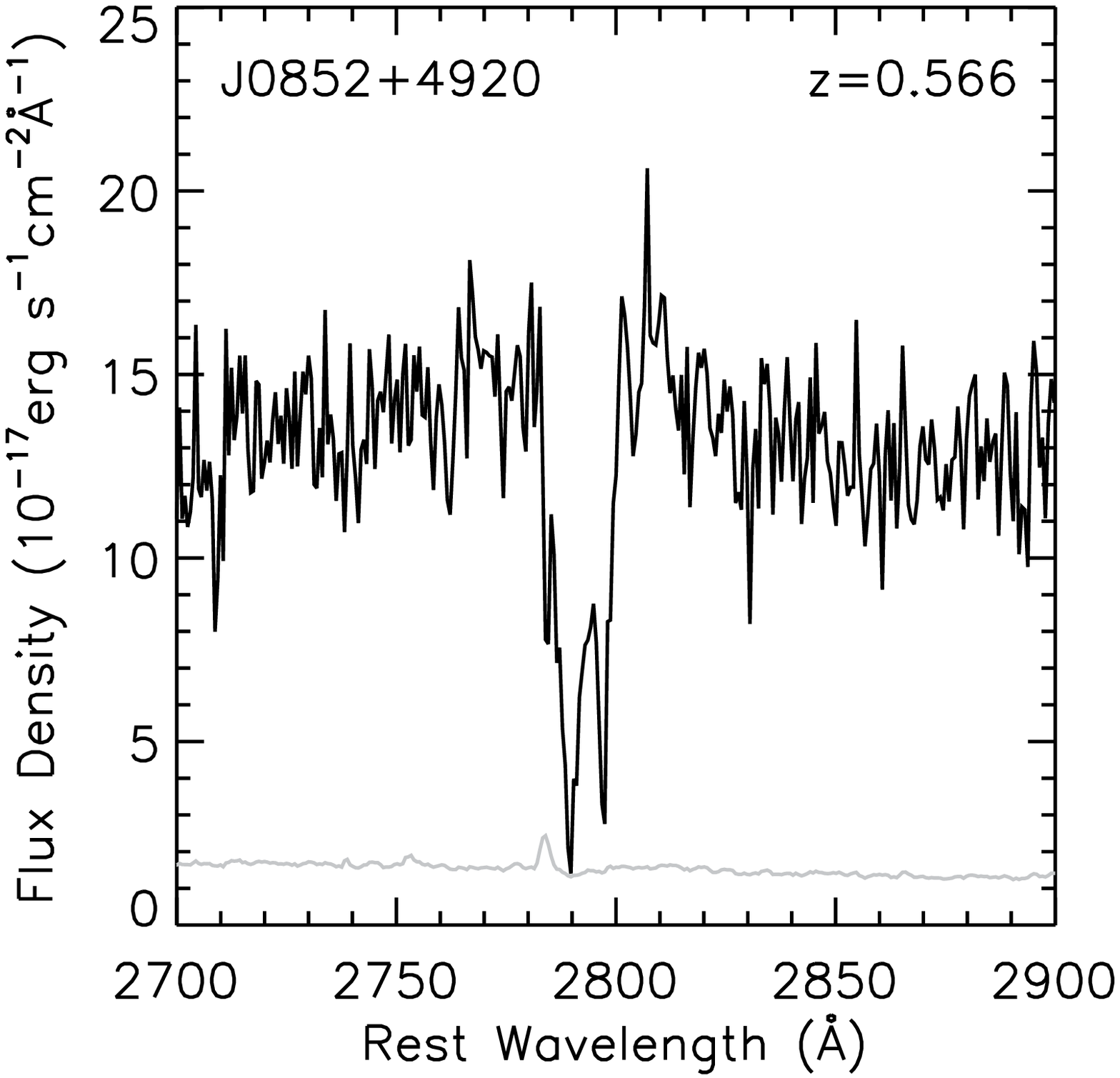}
 \includegraphics[width=0.3\textwidth]{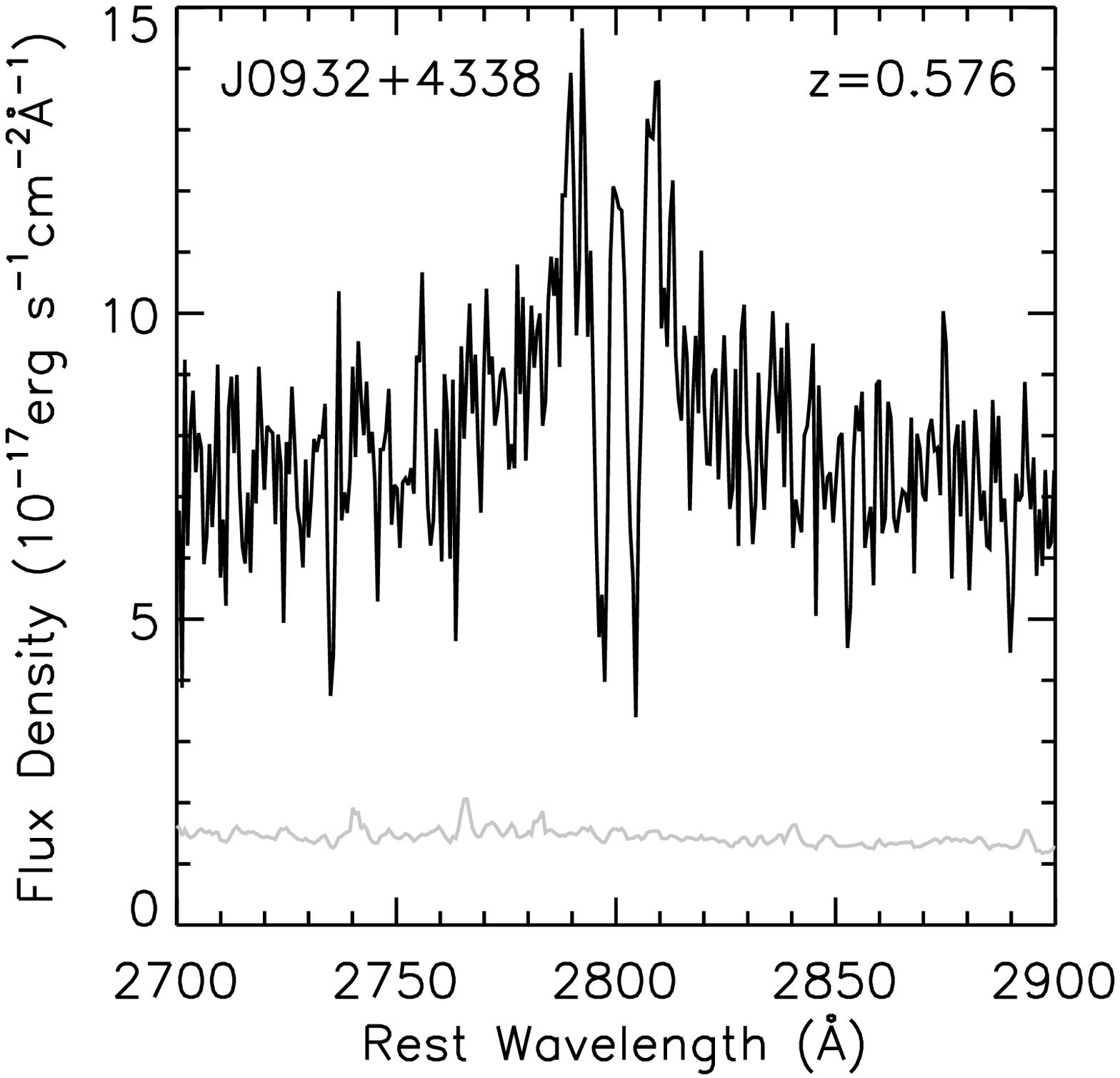}
 \includegraphics[width=0.3\textwidth]{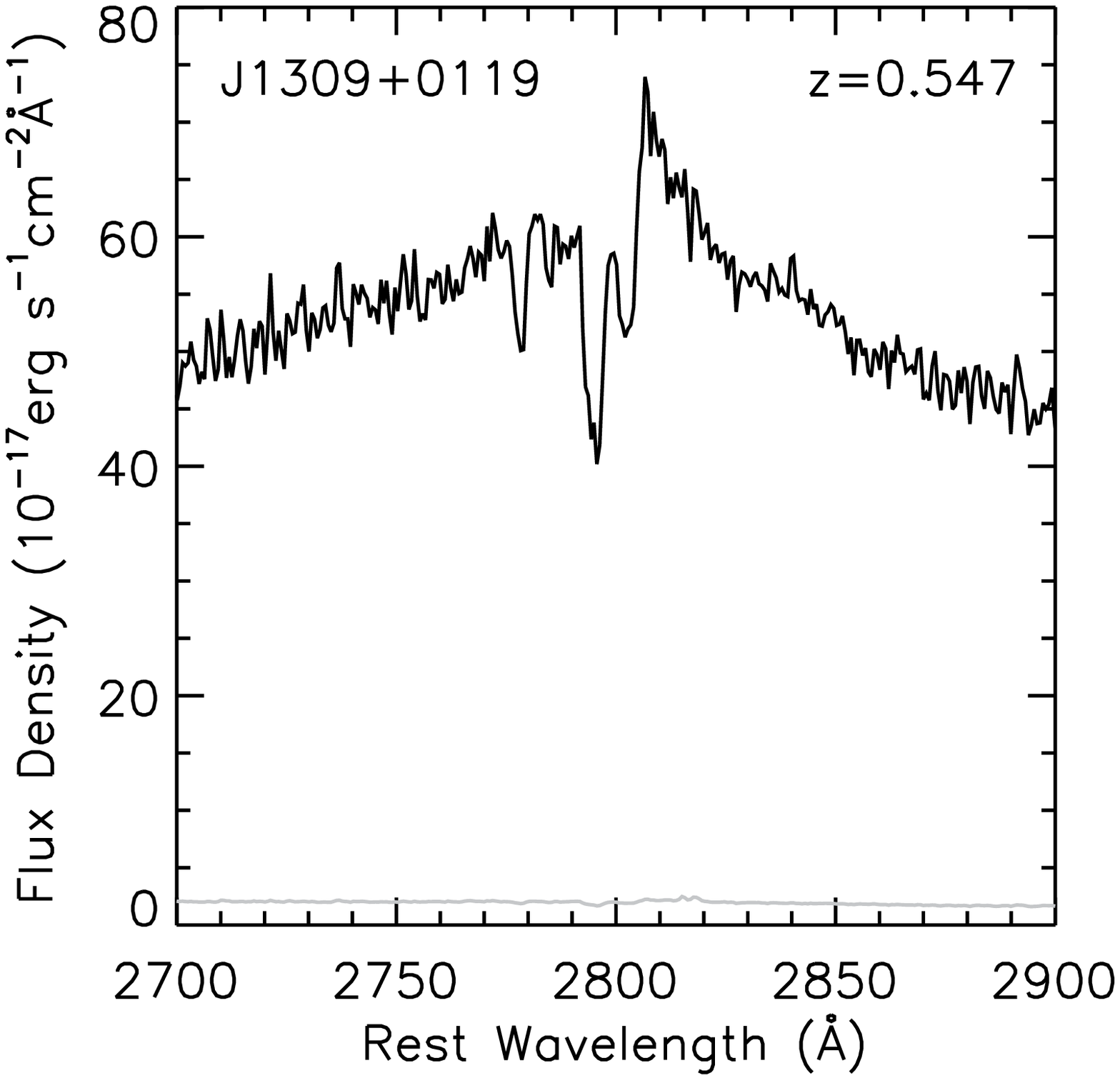}
    \caption{Three AAL quasars in our sample with deep {\em HST} images. {\em Upper:} WFC3-IR images with asinh
    scaling. From left to right: J0852$+$4920 (F125W), J0932$+$4338 (F110W) and
    J1309$+$0119 (F125W). The quasars are located near the centers of the
    images. {\em Bottom:} The \MgII\ region of the optical spectrum from SDSS for the
three objects.}\label{fig:examp}
\end{figure*}


The distinct host properties of AAL quasars in terms of excess \OII\ emission
(with absorber $v_{\rm off}<1500\,{\rm km\,s^{-1}}$) suggest that a
significant fraction of these associated absorbers must be physically
associated with the quasar and its host, and are not from disconnected
external galaxies\footnote{Although these external galaxies may still be
affected by the ionizing flux from the quasar.}. Because the properties of
the active SMBH are similar for AAL quasars and for normal quasars, we expect
that the amount of \OII\ emission contributed by the quasar itself should be
similar as well. We thus interpret the excess \OII\ emission as due to
enhanced star formation within the AAL quasar host galaxy, as \OII\ is a good
indicator of star formation rate in regions with low ionization parameters,
and has been applied for quasar hosts \citep[e.g.,][]{Ho_2005} after
subtracting the \OII\ contribution from the quasar itself. Interestingly,
\cite{Menard_etal_2011} showed a similar connection between \MgII\ absorbers
and star formation in intervening galaxy absorbers. The measured excess of
\OII\ luminosity of the AAL quasar hosts is found to be $L=1.9$ and
$1.2\times10^{41}\, {\rm erg\,s}^{-1}$ for the blueshifted and redshifted
samples respectively. Following \cite{Menard_etal_2011}, we convert these
values to star formation rates using the average scaling coefficient for
galaxies with $0.75<z<1.45$ given by \cite{2009ApJ...701...86Z}. We then find
AAL quasar hosts to have a median excess star formation rate of $\sim$7 (5)
$M_\odot$ yr$^{-1}$ for the blueshifted (redshifted) sample.

A less likely interpretation for the \OII\ excess would invoke different ISM
conditions or element abundance for AAL quasar hosts. Either way, our results
show that AAL quasars represent a distinct population and may indicate a
special evolutionary phase. The clear absence of \OII\ emission at the
redshift of the associated absorbers in the composite spectra disfavors major
mergers (in which case enhanced star formation for the companion would also
have been seen). Our results thus imply that associated \MgII\ absorbers
could either be minor mergers, or galactic to halo-scale gas outflows and
infall associated with the quasar host galaxy. We still expect some AALs from
the clustering of galaxies around quasars \citep[e.g.,][]{Wild_etal_2008}.
Quantifying their contribution requires better knowledge on the strength of
the galaxy-quasar cross-correlation function and pairwise velocity dispersion
(for the specific type of absorbers), as well as the size of the quasar
proximity zone \citep[e.g.,][]{Hennawi_2007}, and is beyond the scope of this
paper.

A number of studies have revealed that dust-reddened quasars have different
properties from classical quasars: heavily reddened (with $E(B-V)\ga 0.5$)
type 1 quasars, selected in the near infrared and radio, usually show strong
\OII\ emission (e.g., Glikman et al.\ 2007). In addition, these systems
display a large fraction (approaching unity) of disturbed morphologies and
interaction (e.g., Urrutia et al.\ 2008). Dust reddened quasars from the SDSS
(with $E(B-V)\sim 0.1$) also show an excess of \OII\ emission compared to
classical quasars (Richards et al.\ 2003). This is in line with the Sanders
\& Mirabel picture that dust-reddened quasars are in the early stage of
evolution after the merger and have higher star formation rate and more dust
obscuration in the nuclei. It is interesting to note that the reddest quasars
also show a higher occurrence of associated absorbers than normal quasars
\citep[e.g.,][]{Aldcroft_etal_1994,Richards_etal_2003,Hopkins_etal_2004}, but
the bulk of AAL quasars have much bluer colors (Fig.\ 3).

AAL quasars thus have properties intermediate between dust-reddened and
classical quasars, in terms of both \OII\ emission and dust reddening. These
findings are consistent with AAL quasars being more evolved than the
dust-reddened quasars following a gas-rich merger. The obscuring material in
the immediate circumnuclear region has largely been dispersed, but the host
star formation rate is still higher than normal quasars. Alternatively, AAL
quasars could be similar to dust-reddened quasars, but are seen along a less
obscured line of sight. In either scenario, we expect AAL quasars to show a
higher fraction of less-developed bulges, disturbed morphologies and
interactions than classical quasars.

An archival search in the {\em HST} database resulted in three AAL quasars in
our sample that have deep ($>10\,$min) imaging data at $z\sim 0.6$, all
observed in serendipitous programs and the data is publicly available. Two of
them, SDSSJ085215.65$+$492040.8 and SDSSJ130952.89$+$011950.6, were observed
with WFC3-IR ($\sim 17$ min and $\sim 13$ min) and WFC3-UVIS ($\sim 28$ min
and $\sim 25$ min). The third object, SDSSJ093210.96$+$433813.1, was observed
with WFC3-IR with a total exposure time of $\sim 40$\,min. In Fig.\
\ref{fig:examp} we show the retrieved WFC3-IR images for these three objects,
and their SDSS spectra around the \MgII\ region. J0932 and J1309 show clear
tidal features and disturbed host morphologies even before subtraction of the
quasar light, while J0852 appears to be in a small interacting group where
the immediate neighbor to the lower-right has a SDSS photometric redshift
consistent with the quasar redshift. Sluse \etal\ (2007) reported a
serendipitously discovered AAL system in the gravitationally lensed quasar
RXS J1131-1231, which also shows evidence of interactions and disturbed
morphology in the reconstructed host image \citep{Claeskens_etal_2006}. Thus
the frequency of disturbed morphologies and close companions of AAL quasars
is much higher than evolved quasars
\citep[e.g.,][]{Bahcall_etal_1997,Dunlop_etal_2003,Veilleux_etal_2009}, and
is more in line with the results for heavily dust-reddened quasars
\citep[e.g.,][]{Urrutia_etal_2008}. This is fully consistent with our
evolutionary scenario, but more data are needed to draw firm conclusions. We
defer a detailed analysis of the {\em HST} data and follow-up studies of
these systems to future work.

Even though our results suggest an intrinsic origin for the bulk of the AALs,
this is a statistical statement and there are still cases where the AALs are
of external origins. Moreover, the nature of these intrinsic AALs is still
unknown. Nevertheless, they provide a unique window to probe the physical
conditions and dynamical processes during the evolution of quasars and their
massive hosts, and deserve in-depth studies, either in a statistical manner
or for individual systems. More than half of these associated absorbers show
blueshifted velocity, among which some could be feedback (either stellar- or
quasar-driven) at work, and possibly leading to a suppression of the host
star formation. The typical velocity of these absorbers (i.e., a few hundred
${\rm km\,s^{-1}}$) is much smaller than those seen in BALs and intrinsic
high velocity narrow absorption lines (NALs), hence these AAL absorbers are
likely to be on scales much larger than nuclear scales\footnote{Associated
absorptions are therefore expected to show little variability on multi-year
timescales compared with BALs and high-velocity intrinsic NALs, which can be
tested with multi-epoch spectroscopy of these absorption line systems (Shen
et al., in preparation).}, and thus are more likely to impact the host
galaxy. Gravitational inflows may also be present during the early stage of
quasar and host formation, either in the form of accreted gas or as part of
an infalling galaxy companion. While intrinsic absorbers with infalling
velocities greater than several hundred ${\rm km\,s^{-1}}$ are likely to be
in the vicinity of the BH, the nature of the more slowly moving absorbers
needs to be explored. In particular, infalling absorbers are expected in
galaxy formation models \citep[e.g.,][]{Keres_etal_2005,Dekel_etal_2009} but
almost never seen in galaxy spectra (Steidel et al. 2010). The possibility
that some of these infalling absorbers might be associated with cold flows
will be explored in future work.




\acknowledgements We thank the anonymous referee for suggestions that led to
improvement of the manuscript. We also thank Michael Strauss, Tim Heckman,
Gordon Richards, Norm Murray, Martin Elvis, and Yuval Birnboim for useful
discussions. YS acknowledges support from the Smithsonian Astrophysical
Observatory (SAO) through a Clay Postdoctoral Fellowship.

Funding for the SDSS and SDSS-II has been provided by the Alfred P. Sloan
Foundation, the Participating Institutions, the National Science Foundation,
the U.S. Department of Energy, the National Aeronautics and Space
Administration, the Japanese Monbukagakusho, the Max Planck Society, and the
Higher Education Funding Council for England. The SDSS Web Site is
http://www.sdss.org/.

\end{document}